\documentclass[preprints,article,accept,moreauthors,pdftex]{mdpi}
\firstpage{1} 
\pubvolume{xx}
\issuenum{1}
\articlenumber{5}
\pubyear{2019}
\copyrightyear{2019}
\history{}
\updates{yes} 

\usepackage[english]{babel}
\usepackage[mathscr]{eucal}
\usepackage{amsbsy,amsfonts,amssymb,makeidx,bm,wasysym}
\usepackage{courier,vmargin,wrapfig,subcaption}

\def\HH{\mathcal{H}}
\def\wm{\overline{w}}
\def\CC{\lambda}
\def\Qm{\overline{\CC}}
\def\BB{\mathscr{B}}
\def\MM{\mathscr{M}}
\def\ma{M_{\mbox{{\tiny P}}}}
\def\Om{\Omega}
\def\fm{\ff_{\star}}
\def\UU{U}
\def\Ep{E_{+}}
\def\tp{\tau_{+}}
\def\ts{\tau_{\star}}
\def\ti{\tau_{0}}
\def\tf{\tau_{1}}
\def\fin{\ff_{0}}
\def\ffi{\ff_{1}}
\def\pf{p_{\ff}}
\def\rf{\rho_{\ff}}
\def\wf{w_{\ff}}
\def\ff{\phi}
\def\kk{k}
\def\x{\textbf{x}}
\def\SS{\mathcal{S}}
\def\R{\mathbb{R}}
\def\geqs{\geqslant}
\def\leqs{\leqslant}
\def\de{\partial}
\def\dd{\displaystyle}

\Title{On the Necessity of Phantom Fields for Solving the Horizon Problem in Scalar Cosmologies}

\Author{Davide Fermi\orcidA{} $^{1,2,}$*$^{,\dagger}$, Massimo Gengo $^{1,\dagger}$ and Livio Pizzocchero\orcidB{} $^{1,2,\dagger}$} 

\AuthorNames{Davide Fermi, Livio Pizzocchero}

\address{%
$^{1}$ \quad Dipartimento di Matematica, Universit\`a di Milano, Via C. Saldini 50, I-20133 Milano, Italy; mandi8911@gmail.com (M.G.); livio.pizzocchero@unimi.it (L.P.)\\
$^{2}$ \quad Istituto Nazionale di Fisica Nucleare, Sezione di Milano, I-20133 Milano, Italy}
\corres{Correspondence: davide.fermi@unimi.it}

\firstnote{These authors contributed equally to this work.}

\abstract{We discuss the particle horizon problem in the framework of spatially homogeneous and isotropic scalar cosmologies. To this purpose we consider a Friedmann--Lema\^{i}tre--Robertson--Walker (FLRW) spacetime with possibly non-zero spatial sectional curvature (and arbitrary dimension), and assume that the content
of the universe is a family of perfect fluids, plus a scalar field that can be a quintessence or a phantom (depending on the sign of the kinetic part in its action functional). We show that the occurrence of a particle horizon is unavoidable if the field is a quintessence, the spatial curvature is non-positive and the usual energy conditions are fulfilled by the perfect fluids. As a partial converse, we present three solvable models where a phantom is present in addition to a perfect fluid, and no particle horizon~appears.}




\begin{document}

\section{Introduction}
There are many reasons for considering cosmological models where
a classical scalar field is present, in addition to ordinary
matter and radiation.

Since the beginning of 1980s, it was suggested that a scalar
field could be related the alleged inflationary behavior of the
early universe~\cite{Guth,Linde,Linde2,MaCo}. The subsequent
literature is too wide for an exhaustive account,
thus we limit ourselves to mentioning  
\cite{Ba1,Coley,GuthRep,Linde4,Linde5,Lucc,Olive}.

In the late 1980s, a scalar field was proposed as a simple,
but nontrivial dynamical model for dark energy~\cite{Peeb}.
The presence of some kind of dark energy was just
a possible scenario in this seminal work, but became a
consolidated idea after the experimental discovery
of the present time accelerated expansion of the universe
\cite{Perl,Riess}. Even the literature on scalar models
of dark energy is enormous; 
here we only cite~\cite{CaldDE,Eli04,Mat,Pie,Sah,Star,Sol}.

Starting from the above physical motivations,
many authors focused the attention on
the mathematically rigorous treatment of cosmological models
with scalar fields. Exact solutions of the Einstein
equations for these models are obtained in
\cite{Ba0,Burd,Cat,Chim,TG,Ma1,Ma2,Eas,Fre,Peeb,PTB14,PTB15,Pie,Rub}.
Again, in the area of exact solutions,
we would mention a peculiar, ``inverse'' approach
to the evolution equations of scalar cosmologies: here
the time dependence of the scale factor (or
of some other physically relevant quantity) is
prescribed, and the Einstein equations are used
to determine the other observables of
the model, including the self-interaction
potential of the scalar field~\cite{Ba1,BaPa,ElMa,Lucc}.

Up to now, we have generically employed
the term ``scalar field'' to indicate
either a \emph{canonical} field (also
called a \emph{quintessence} in cosmology), or a
\emph{phantom} field, with an anomalous
sign in the kinetic part of
the action functional. 

Phantom fields were originally introduced
in cosmology as toy models in which
the equation of state parameter (i.e.,
the ratio of the pressure to the density)
can attain values below $-1$~\cite{Cald, Car};
it~is perhaps superfluous to recall that
such a behavior implies a violation of
the weak energy condition~\cite{Hawk}.
It is well known that the
(renormalized) vacuum expectation value of
the stress--energy tensor of
a quantized canonical field (see, e.g.,~\cite{BCME,FP})
also violates this energy condition; thus,
phantom fields can be viewed as
simplified models for quantum vacua
of canonical fields, as first remarked in~\cite{OdNo03}.

In applications to cosmology, phantom
fields were mainly considered in connection
with the present day accelerated expansion
\cite{Cald,Cap,Car,Dutta,Eli04,Gibb,Sin}.
The aim of the present work is to provide a
different motivation to focus the attention
on phantoms, connected with the early
history of universe. The subject that we
wish to link to phantom fields is
the well known horizon problem, which
can be described as follows for any
Friedmann--Lema\^{i}tre--Robertson--Walker (FLRW)
cosmology~\cite{ElSt,Hawk,PlKr,Rind,Wald}.
Depending on the behavior of the scale factor close to
the Big Bang, at a given instant,
there are two alternatives for
any particle at rest in the FLRW
frame: the spatial region that has already
interacted with the particle either is confined within
a suitable horizon, or~is the whole
space (no particle horizon). Of course,
the absence of a horizon is the physically
most pleasant case: in fact, due to
the causal connection of the whole space
to any point, we can invoke
some kind of thermalization process
to explain the homogeneity of the
universe.

In this paper, we discuss the horizon problem for 
FLRW cosmologies with a scalar field and a finite number of
perfect fluids, each of them
representing some (reasonable) kind of matter
or radiation. Here are the main outputs
of this analysis:

\begin{enumerate}[align=parleft,leftmargin=*,labelsep=5.5mm]
\item[(I)]
Putting the usual energy conditions on the state 
equations of the fluids and assuming
a non-positive spatial curvature, we prove
that a \emph{quintessence} (canonical scalar field)
with an \emph{arbitrary} self-interaction potential
\emph{always produces a particle horizon}. This fact
was probably noticed in specific cases but, to
the best of our knowledge, it was never proved
with the present generality.
The conditions on the state equations of the fluids
that we assume are satisfied, e.g., by dust and by a radiation gas.
Our requirement of non-positive curvature is,
mainly, to simplify the analysis;
we hope to discuss elsewhere the case
of positive curvature, using~arguments
similar to the present ones.
\vspace{0.2cm}
\item[(II)] In view of (I), the quest for scalar
cosmologies with (non-positive curvature and)
no particle horizon naturally brings us to
consider the phantom case. We show that
a \emph{phantom} field with a suitable self-potential,
in presence of some reasonable kind of perfect fluid,
can indeed produce a FLRW cosmology with no horizon:
three examples of this kind are presented.
In~the first example, we have a spatially flat
de Sitter universe (with a Big Bang at
an infinitely remote instant); the perfect
fluid has a quite arbitrary equation of state,
and the phantom field has a quadratic
self-interaction potential. In the second
example, the spatial curvature is negative,
the Big Bang occurs at a finite time,
the perfect fluid is a radiation gas
and the phantom has a quartic, self-interaction
potential of Higgs  type. In the third example, the~spatial
curvature is zero, the Big Bang happens at a
finite time, a radiation gas is present
and the phantom has an           {ad hoc} self-potential,
not expressible via elementary functions.
All~these examples present an accelerated expansion,
and an exponential divergence of the
scale factor over long times.
\end{enumerate}

Concerning Item (II), we are aware that phantom fields are
potentially responsible for a number of pathological features,
such as a Big Rip (the simultaneous divergence of the scale
factor, and of the density and pressure at a finite time after
the Big Bang) or a Little Rip (divergence of the same quantities
in the infinitely far future)~\cite{Odi1,Bre11,CalW,LRip,LRip2,OdNo05};
on top of that, it was pointed out that the thermodynamics of phantom fields
may cause the appearance of negative temperatures or of a negative
total entropy for the universe~\cite{Bre04,Gon04,Myu09,OdNo04,OdNo05b}.
Discussing these issues is beyond the purposes of the present work
{\footnote{Another pathology of phantoms, namely, their vacuum instability
at the quantum level~\cite{Car,Cli,Cop}, is not   considered here:
our phantom field is a purely classical object, perhaps acceptable
as a toy model of a quantized \emph{canonical} field for the
reasons already mentioned when we cited~\cite{OdNo03}.}};
here, we are led to consider a phantom by purely logical arguments
related to Proposition~\ref{prop}, which are inescapable if one
wants a cosmological model with no horizon and matter
fulfilling the usual energy conditions. 
In any case, in the examples of this paper the phantom field
never produces a Big Rip nor a Little Rip.

Let us briefly illustrate the organization of the paper.
Most of the paper refers to a spacetime of
arbitrary dimension $d+1$, where $d \geqs 2$; of course,
$d=3$ is the most physically relevant case.
Sections~\ref{secgen} and~\ref{secCos}
review some basic facts on
gravitation in presence of scalar fields and perfect
fluids, and on FLRW cosmologies based on the same actors
(including the related energy conditions); 
the horizon problem is summarized in Section
\ref{secHor}. The aim of these sections
is to fix some standards, and to present
the Einstein (and Klein--Gordon) equations
in a form convenient for our subsequent
considerations.
Section~\ref{secQuin} treats the subject
indicated in the previous Item (I), i.e.,
the fact that a canonical scalar field
always produces a particle horizon
(in the case of non-positive spatial
curvature, and with minimal assumptions
on the perfect fluids); the general
formulation of this result is
contained in Proposition~\ref{prop}.
Section~\ref{secPhan} has the content described
in the previous Item (II), i.e., it presents
three cosmologies with a phantom field,
a perfect fluid and no particle horizon; these
are described in Sections~\ref{subs1}--\ref{subs3}, respectively.
\vspace{0.1cm}

{\textbf{A final remark.} One of the Reviewers of this paper
pointed out that a simpler solution of the horizon problem,
avoiding the use of phantom fields, would be to consider a universe
filled with a canonical scalar field violating the strong energy condition,
at least in an initial interval. 
However, the aim of this paper is to discuss the horizon problem
in a framework where the content of the universe \emph{includes some
type of ordinary matter}, \emph{fulfilling
the usual energy conditions}; according to our Proposition~\ref{prop},
it is just the joint presence of such ordinary matter and of
a canonical scalar field that produces a particle horizon.
This is the reason, besides ordinary matter,
we are led to consider a phantom field.}

\section{Generalities on Gravity, Perfect Fluids and Scalar Fields}\label{secgen}
Throughout the paper, we work in natural units, in which the speed
of light and the reduced Planck constant are $c = 1$ and $\hbar = 1$.

As a starting point, let us consider a general model
living in a $(d+1)$-dimensional spacetime, of spatial dimension $d \geqs 2$.
Let $x = (x^\mu)_{\mu = 0,...,d}$ be any set of spacetime coordinates.
The metric $g_{\mu\nu}$ has signature $(-,+,...,+)$; the corresponding
covariant derivative, Ricci tensor and scalar curvature are indicated,
respectively, with $\nabla_{\!\mu}$\,, $R_{\mu\nu}$ and $R$.
We assume the model to include the~following:

\begin{itemize}[leftmargin=2.3em,labelsep=5.5mm]
\item {$n \geqs 1$} species of non-interacting perfect fluids, all with the same $(d+1)$-velocity.
For any $i \in \{1,...,n\}$, we suppose that the mass-energy density
$\rho_i$ and the pressure $p_i$ of the $i$th fluid are related by
a barotropic equation of state of the form $p_i = p_i(\rho_i)$. 

\item A real classical scalar field $\ff$, minimally coupled to gravity
and self-interacting with potential~$V(\ff)$.
\end{itemize}

The action functional governing the dynamics of the system
outlined above is
\begin{equation}\label{act}
\SS \,:= \int d^{d+1}\!x\; \sqrt{-g} \left[
{R \over 2\kappa_d}
- \sum_{i = 1}^n \rho_i
- {\sigma \over 2}\, g^{\mu\nu} \de_\mu \ff\, \de_\nu \ff - V(\ff) \right],
\end{equation}
where $g$ denotes the determinant of the metric $g_{\mu\nu}$ and
$\kappa_d$ is the Einstein gravitational constant in $d+1$ dimensions.
Here and in the sequel, $\sigma$ is a pure sign parameter:
in compliance with the standard nomenclature, we   call the field $\ff$
a \emph{quintessence} if $\sigma = +1$ and a \emph{phantom} if $\sigma = -1$.

The evolution equations for the model can be derived requiring the action
$\SS$ to be stationary under variations of the metric, of the fluids' histories
and of the field.

Firstly, the stationarity condition for variations of the metric $g_{\mu\nu}$ yields
the Einstein equations
\begin{equation}\label{EinstEq}
R_{\mu\nu} - {1 \over 2}\,R\,g_{\mu\nu} =
\kappa_d \left(\,\sum_{i = 1}^n T^{i}_{\mu\nu} + T^{\ff}_{\mu \nu}\right);
\end{equation}
these involve the stress--energy tensors of the fluids and of the field,
defined, respectively, as
\begin{gather}
T^{i}_{\mu\nu} := (p_i + \rho_i)\, U_\mu\, U_\nu + \,p_i\, g_{\mu\nu} \quad\;
\mbox{for\; $i \in \{1,...,n\}$}\,, \label{Tmdef} \\
T^{\ff}_{\mu\nu} := \sigma\! \left(\de_\mu \ff\,\de_\nu \ff
- {1 \over 2}\,g_{\mu\nu}\,\de^\gamma \ff\, \de_\gamma\ff\right) - V(\ff)\,g_{\mu\nu}\, , \label{Tfdef}
\end{gather}
with $U^\mu$ indicating the common $(d+1)$-velocity vector field of all the fluids.

Secondly, stationarity of $\SS$ under variations of the fluids' histories
(defined as in~(\cite{Hawk},~Section~3.3))
leads to the separate conservation laws
\begin{equation}\label{Tmcon}
\nabla^{\mu\,} T^{i}_{\mu\nu} = 0 \quad\; \mbox{for\; $i \in \{1,...,n\}$}\,;
\end{equation}
taking this into account and using the contracted Bianchi identity for the
Einstein equations 
\eqref{EinstEq}, we also obtain the analogous conservation law
\begin{equation}\label{Tfcon}
\nabla^{\mu\,} T^{\ff}_{\mu\nu} = 0\,.
\end{equation}

Lastly, the stationarity condition for $\SS$ with respect to variations
of the field $\ff$ gives rise to the Klein--Gordon equation
\begin{equation}\label{KG}
-\,\sigma\, \nabla_{\!\mu}\nabla^{\mu} \ff + V'(\ff) = 0\,.
\end{equation}

As well known, in the phantom case $\sigma = -1$, the evolution equation 
\eqref{KG} implies an exotic behavior of the field $\ff$, which spontaneously evolves
towards maxima (rather than minima) of the potential $V(\ff)$.

On the other hand, notice that from Equation           \eqref{Tfdef} it follows
$\nabla^{\mu\,} T^{\ff}_{\mu\nu} = (\sigma \nabla_{\!\mu} \nabla^\mu \ff - V'(\ff))\,\de_\nu \ff$,
for both $\sigma = +1$ and $\sigma = -1$; thus, whenever $\de_\nu \ff \neq 0$,
the conservation law 
\eqref{Tfcon} and the field equation 
\eqref{KG} are
indeed equivalent. We have already noted that the Einstein equations 
\eqref{EinstEq} and the conservation laws 
\eqref{Tmcon} for the matter fluids
imply Equation           \eqref{Tfcon}, thus they also convey the Klein--Gordon equation 
\eqref{KG} whenever $\de_\nu \ff \neq 0$.

\section{The Reference Cosmological Model}\label{secCos}
The general framework of the previous section can be used
to set up a cosmological model, where the perfect fluids
are meant to provide an effective description of different
types of matter or radiation; here, the scalar field models
dark energy over long times, and possibly triggers
an initial inflationary behavior.

As customary, we~construct a spatially homogeneous and isotropic model.
Accordingly, we~consider a $(d+1)$-dimensional FLRW
spacetime ($d \geqs 2$), which is the product of a time interval
with its natural coordinate $\tau$ (the ``cosmic time'') and
of a $d$-dimensional complete, simply~connected Riemannian manifold
$\MM_\kk$ (the ``space''), of constant sectional curvature $\kk$
(i.e, a flat Euclidean space if $\kk=0$, a hyperbolic space
if $\kk<0$ and a spherical surface if $\kk>0$); we write
$d \ell^2_\kk$ for the line element of $\MM_\kk$ and often
refer to $\kk$ as to the ``spatial curvature''.
The spacetime line element is
\begin{equation}\label{dsa}
ds^2 = -\,d\tau^2\! + a(\tau)^2\,d\ell_\kk^2\,,
\end{equation}
where $a(\tau)>0$ is the dimensionless scale factor.

Concerning the $n$ species of perfect fluids, for any $i \in \{1,...,n\}$,
we postulate the linear equation of state
\begin{equation}\label{wi}
p_i = w_i\,\rho_i\,, \qquad \mbox{with constant $w_i \in \R$}\,.
\end{equation}

Besides, we assume these fluids to be co-moving with the FLRW
frame, so that the \linebreak  $(d+1)$-velocity $U^\mu$ appearing in the stress--energy
tensors in Equation \eqref{Tmdef} coincides with the coordinate vector field $\de_\tau$.
In agreement with the homogeneity hypothesis, we further suppose that their
mass--energy densities and pressures depend only on cosmic time, i.e.,
\begin{equation}\label{rhop}
\rho_i \equiv \rho_i(\tau)\,, \qquad p_i \equiv p_i(\tau)\,.
\end{equation}

On account of the assumptions stated above, the conservation equations 
\eqref{Tmcon} reduce to
\begin{equation}\label{cont}
\dot{\rho}_i + d\, (1 + w_i)\,{\dot{a} \over a}\,\rho_i = 0
\quad\; \mbox{for\; $i \in \{1,...,n\}$}
\end{equation}
(here and in the sequel, dots denote derivatives with respect to $\tau$;
e.g., $\dot{a} \equiv da/d\tau$).
The differential equations 
\eqref{cont} are solved by the redshift relations
\begin{equation}\label{eqred}
\rho_i(\tau) = {d\,(d-1) \over 2\,\kappa_d}\;{\CC_i \over a(\tau)^{d(1 + w_i)}}\;,
\end{equation}
where $\CC_i$ is a constant. The factor $d(d\!-\!1)/(2 \kappa_d)$ has been
introduced in Equation           \eqref{eqred} for future convenience; note that
$d(d\!-\!1)\CC_i/(2 \kappa_d)$ is the density of the $i$th fluid
at the instant when the scale factor is $a(\tau) = 1$~{\footnote{The dimension of $\CC_i$ is length$^{-2}$ $=$ time$^{-2}$
in our units with $c = 1$, $\hbar = 1$.}}.

Next, let us consider the field $\ff$; to comply with the homogeneity
hypothesis, we assume that also $\ff$ only depends on cosmic time, i.e.,
\vspace{-0.1cm}
\begin{equation}\label{fft}
\ff \equiv \ff(\tau)\,.
\end{equation}
In this case, the stress--energy tensor 
\eqref{Tfdef} is also of perfect fluid form
\begin{equation}\label{Tfflu}
T^{\ff}_{\mu\nu}\! = (\pf + \rf)\, U_\mu\, U_\nu + \pf\, g_{\mu\nu}\,,
\end{equation}
with density $\rf$ and pressure $\pf$, respectively, given by
\begin{equation}\label{rhofpf}
\rf := {\sigma \over 2}\,\dot{\ff}^2 + V(\ff)\, , \qquad
\pf := {\sigma \over 2}\,\dot{\ff}^2 - V(\ff)\, .
\end{equation}

The equation of state parameter for the scalar
field can be defined as follows, whenever~ $\rf \neq 0$:
\begin{equation}\label{wff}
\wf := {\pf \over \rf} =
{\sigma\,(\dot{\ff}^2\!/2) - V(\ff) \over \sigma\,(\dot{\ff}^2\!/2) + V(\ff)}\,.
\end{equation}
Note that $\wf$ depends on time (with a law determined by the
solution of the cosmological model), while the analogous
parameters $w_i$ ($i \in \{1,...,n\}$) of the perfect fluids mentioned
before are universal constants; this is a good reason to
distinguish between any of these $n$ fluids and the scalar field{\footnote{On the contrary, admitting equations of state with
variable parameters, one could strengthen the analogies between
perfect fluids and homogeneous scalar fields in FLRW cosmologies~\cite{Odi1}.}}.

Taking into account all the facts mentioned above,
the Einstein equations 
\eqref{EinstEq} ~become~{\footnote{To be precise, let us denote with $(x^j)_{j \,=\, 1,...,d}$
any coordinate system for $\MM_{\kk}$ (see the comments before Equation           \eqref{dsa}),
and~let us refer to the spacetime coordinates $(x^\mu)_{\mu \,=\, 0,1,...,d}$
where $x^0 := \tau$. Then, the Einstein equations 
\eqref{EinstEq}
(with the expressions 
\eqref{eqred} for the densities $\rho_i$
and 
\eqref{wi} for the pressures $p_i$) are equivalent, respectively,
to Equation           \eqref{E1} for $\mu=\nu=0$ and to Equation           \eqref{E2} for $\mu,\nu \in \{1,...,d\}$;
in the mixed cases $\mu=0$, $\nu \in \{1,...,d\}$ or
$\mu \in \{1,...d\}$, $\nu=0$, the Einstein equations are
trivially satisfied. See, e.g., Reference~\cite{TG} for more details
on this computation and on other statements in this
section.}}
\begin{gather}
{d\,(d - 1) \over 2} \left({\dot{a}^2 \over a^2} + {\kk \over a^2}
- \sum_{i = 1}^n {\CC_i \over a^{d (1 + w_i)}} \right)
- \kappa_d \Bigg({\sigma \over 2}\,\dot{\ff}^2 + V(\ff) \!\Bigg) =\, 0\,, \label{E1}\\
(d - 1)\!\left({\ddot{a} \over a} + {d - 2 \over 2}\,{\dot{a}^2 \over a^2}
+ {d - 2 \over 2}\,{\kk \over a^2}
+ {d \over 2} \sum_{i = 1}^n {w_i\,\CC_i \over a^{d (1 + w_i)}} \right)
+ \kappa_d \Bigg({\sigma \over 2}\,\dot{\ff}^2 - V(\ff) \!\Bigg) =\, 0\,. \label{E2}
\end{gather}

To proceed, let us remark that, in the present framework,
the Klein--Gordon equation 
\eqref{KG} reduces to
\begin{equation}\label{ffeq}
\ddot{\ff} + d \,{\dot{a} \over a}\,\dot{\ff} + \sigma\, V'(\ff) = 0\,.
\end{equation}

We already mentioned at the end of Section~\ref{secgen} that,
for the general models described therein, the Klein--Gordon equation
is indeed a consequence of the Einstein equations whenever the scalar
field has non-zero gradient. In the present homogeneous and isotropic
framework, this fact can be ascribed to the identity
\begin{equation} \label{ident}
{d \over d\tau} \Big[a^d \cdot \eqref{E1} \Big] - d\,a^{d-1}\,\dot{a} \cdot \eqref{E2}
+ \sigma\,\kappa_d\,a^d \, \dot{\ff} \cdot \eqref{ffeq} \,=\, 0\;,
\end{equation}
where the notations 
\eqref{E1}--\eqref{ffeq} stand for the left-hand sides of
Equations~\eqref{E1}--\eqref{ffeq}; this makes
patent that the Einstein equations 
\eqref{E1} and \eqref{E2} imply the Klein--Gordon equation 
\eqref{ffeq} at all times $\tau$ such that $\dot{\ff}(\tau) \neq 0$
{\footnote{As a partial converse of this statement, notice the following:
if Equations          \eqref{E2} and \eqref{ffeq} hold true at all times in an interval
and Equation           \eqref{E1} is fulfilled at some instant in this interval,
from the identity 
\eqref{ident} it follows that Equation           \eqref{E1}
holds true at all times in the interval.}}.

For later reference, let us point out that Equations          \eqref{E1} and \eqref{E2} are
equivalent to the pair
\begin{gather}
\dot{\ff}^2 =\, \sigma\,{d-1 \over \kappa_d}
\left[{\dot{a}^2 \over a^2} - {\ddot{a} \over a} + {\kk \over a^2}
- {d \over 2} \sum_{i = 1}^n {(1 + w_i)\,\CC_i \over a^{d (1 + w_i)}}\right] , \label{Fid2}\\
V(\ff) \,=\, {d-1 \over 2\,\kappa_d}
\left[(d-1)\,{\dot{a}^2 \over a^2} + {\ddot{a} \over a} + (d-1)\,{\kk \over a^2}
- {d \over 2}\sum_{i = 1}^n {(1 - w_i)\,\CC_i \over a^{d (1 + w_i)}}\right] . \label{Vf}
\end{gather}

Let us conclude this subsection with two remarks:

\begin{enumerate}[leftmargin=21pt,labelsep=7pt]
\item[$(\alpha)$] There are solutions of
Equations          \eqref{E1}--\eqref{ffeq} with $\ff(\tau) =$ const.\! $\equiv \fm$;
in particular, Equation~\eqref{ffeq} shows that a solution of this kind is possible
if and only if $V'(\fm) = 0$. On the other hand, from Equations          \eqref{rhofpf} and 
\eqref{wff} we infer that a function $\tau \to \ff(\tau)$
has a constant value $\fm$ if and only if $\pf = - \rf$, i.e.,\!
(assuming $\rf \neq 0$), $\wf = -1$ at all times.
If this occurs, Equation           \eqref{Tfflu} for the field stress--energy tensor
gives $T^{\ff}_{\mu \nu} = - \,V(\fm)\, g_{\mu \nu}$;
this is the stress--energy tensor corresponding to a cosmological constant
term in the Einstein equations, thus the scalar field is said to behave
as a cosmological constant.
\item[$(\beta)$] For future reference, it is convenient to review some
known facts about the standard energy conditions (see~\cite{Hawk} for the usual formulation
in spatial dimension $d = 3$ and~\cite{Maeda} for its extension to arbitrary $d \in \{2,3,4,...\}$).
Let us fix $i \in \{1,...,n\}$ and consider the $i$th fluid of the previously mentioned
family; its stress--energy tensor $T^{i}_{\mu \nu}$ (see Equation           \eqref{Tmdef})
fulfils the weak energy condition if and only if
$\rho_{i} \geqs 0$ and $p_{i} + \rho_{i} \geqs 0$, while it fulfills
the strong energy condition if
$p_{i} + \rho_{i} \geqs 0$ and $d\, p_{i} + (d-2)\, \rho_{i} \geqs 0$.
With the assumption $\rho_{i} > 0$ (positive density), the weak and
strong energy conditions are, respectively, equivalent to the relations
\begin{gather}
w_{i} \geqs -1\,, \label{WEC}\\
w_{i} \geqs {2 \over d} - 1\,. \label{SEC}
\end{gather}

Of course, one can make similar statements for the scalar field replacing
$\rho_{i},p_{i},w_{i}$ with $\rf,\pf,\wf$.
\end{enumerate}

\subsection{The Particle Horizon Problem}\label{secHor}
The subject of the present subsection is relevant for any cosmological
model based on a spacetime of the FLRW
type, described by Equation           \eqref{dsa} and by the related comments.
Given any such model, we assume the following:

\begin{enumerate}[align=parleft,leftmargin=*,labelsep=5.5mm]
\item[(i)] The cosmic time $\tau$ ranges in an interval $(\ti, \tf)$,
where $-\infty \leqs \ti < \tf \leqs + \infty$; besides,
the~scale factor $\tau \mapsto a(\tau)>0$ is smooth on this interval.
\vspace{0.15cm}
\item[(ii)] A \emph{Big Bang} occurs at $\ti$, meaning that
\begin{equation}\label{BB}
a(\tau) \to 0 \quad \mbox{for\; $\tau \to \ti$}\,.
\end{equation}
\end{enumerate}

Correspondingly, let us consider the lapse of conformal time that has
passed from the Big Bang up to any cosmic time $\tau \in (\ti, \tf)$, namely,
\begin{equation}\label{Theta}
\Theta(\tau) := \!\int_{\ti}^{\tau} {d\tau' \over a(\tau')} \,\in\, (0,+\infty]\,.
\end{equation}

In our units with $c = 1$, $\Theta(\tau)$ represents the distance
in $\MM_{\kk}$ (with respect to the metric $d \ell^2_\kk$) travelled until cosmic
time $\tau$ by a light signal emitted at the Big Bang and propagating freely.
On~account of this, given any co-moving FLRW
particle with position
$\x$ in $\MM_{\kk}$, the co-moving particles that had enough time
to interact causally with it before time $\tau$ are those with position in the ball
of $\MM_{\kk}$ with center in $\x$ and radius $\Theta(\tau)$, which we
denote with $\BB(\x,\Theta(\tau))$.
If this ball is smaller than the whole space $\MM_\kk$, it has
a boundary $\partial \BB(\x,\Theta(\tau))$, which is referred
to as the \emph{particle horizon} of $\x$ at cosmic time $\tau$
\cite{ElSt,Hawk,PlKr,Rind,Wald}.

Next, let us notice that, for any $\x \in \MM_{\kk}$,
the sup of the distances between $\x$ and the other points
of $\MM_{\kk}$ is independent of $\x$ and equals{\footnote{This statement is obvious if $\kk \leqs 0$,
since $\MM_{\kk}$ is either a Euclidean or a hyperbolic space;
if $\kk >0$, $\MM_{\kk}$ is a spherical surface of radius
$1/\sqrt{\kk}$ and the maximum of the distances from $\x$,
attained at the antipodal point, is half the length of a
great circle.}}
\begin{equation}\label{delk}
\delta_{\kk} := \left\{\!\!\begin{array}{ll}
\dd{\infty} & \dd{\mbox{for $\kk \leqs 0$}\,,} \vspace{0.1cm}\\
\dd{{\pi/\sqrt{\kk}}} & \dd{\mbox{for $\kk > 0$}\,.} \\
\end{array}\right.
\end{equation}

Thus, for any $\x \in \MM_{\kk}$, there exists no particle horizon
at cosmic time $\tau$ if and only if
\begin{equation}\label{horcon}
\Theta(\tau) \geqs \delta_{\kk}\,.
\end{equation}

If this condition is fulfilled, any $\x \in \MM_{\kk}$ interacts
causally with any other point before time $\tau$; thus, the homogeneity
of the universe at $\tau$ can be explained making
reference to classical thermalization processes,
damping out possible initial inhomogeneities.
If the condition 
\eqref{horcon}  is violated, homogeneity
of the universe at time $\tau$ cannot be explained by this mechanism:
this is usually referred to as the ``\emph{horizon problem}''.

For $\kk \leqs 0$, Equations          \eqref{delk} and \eqref{horcon} entail that
there is no particle horizon at a time $\tau$ if and~only~if
\begin{equation}\label{nohor}
\Theta(\tau) = + \infty \,.
\end{equation}

Let us remark that, due to the definition 
\eqref{Theta} of $\Theta(\tau)$
and depending on the way in which $1/a(\tau)$ diverges for $\tau \to \ti$,
we have the following alternatives:
\begin{equation}\label{altern}
\mbox{either\; $\Theta(\tau)\!<\!+\infty$ for all $\tau\!\in\!(\ti,\tf)$\;\;
or\; $\Theta(\tau) = +\infty$ for all $\tau\!\in\!(\ti,\tf)$}\,.
\end{equation}

Addressing the horizon problem for $\kk > 0$
typically demands a precise quantitative analysis,
due to the finiteness of $\delta_\kk$ in Equation           \eqref{horcon}.
Even though the arguments to be presented in the next section
could be in part generalized to settings with $\kk > 0$, in the sequel for simplicity,
we do not consider this case.

\section{Particle Horizon in the Quintessence Case $\sigma = +1$}\label{secQuin}
In this section, we show that a homogeneous and isotropic universe
of non-negative spatial curvature, filled with perfect fluids
and a quintessence, under minimal additional conditions has a
finite particle horizon. Of course, this result rises the
interpretation problems mentioned in~Section~\ref{secHor}.

More precisely, let us consider a cosmological model
as in Section~\ref{secCos}, with the following~features:

\begin{enumerate}[align=parleft,leftmargin=*,labelsep=5.5mm]
\item[(i)] The cosmic time $\tau$ ranges in an interval $(\ti,\tf)$,
where $-\infty \!\leqs\! \ti \!<\! \tf \!\leqs\! + \infty$;
besides, the~functions $\tau \mapsto a(\tau)\!>\!0$, $\tau \mapsto \ff(\tau)$
are smooth on $(\ti,\tf)$, and fulfill Equations          \eqref{E1}--\eqref{ffeq}.
\item[(ii)] There is a Big Bang at $\ti$, namely $a(\tau) \to 0$ for $\tau \to \ti$.

\item[(iii)] The spatial curvature is non-positive, i.e.,
\begin{equation}\label{kneg}
\kk \leqs 0\,. \vspace{0.cm}
\end{equation}
\item[(iv)] The scalar field $\ff$ is a quintessence, i.e.,
\begin{equation}\label{sip}
\sigma = +1\,. \vspace{0.cm}
\end{equation}
\item[(v)] The {$n \geqs 1$} perfect fluids describing
the matter content of the universe are such that
\begin{gather}
\CC_i > 0 \quad \mbox{for all\, $i \in \{1,...,n\}$}\,; \label{Ci} \\
w_i \geqs -1 \quad \mbox{for all\, $i \in \{1,...,n\}$}\,, \qquad
w_i > {2 \over d} - 1 \quad \mbox{for some\, $i \in \{1,...,n\}$}\,. \label{wm}
\end{gather}

Let us recall that $\CC_i$ is the constant coefficient introduced in Equation           \eqref{eqred},
while $w_i$ is the parameter in the equation of state 
\eqref{wi}. The condition 
\eqref{Ci} means that all fluids have positive densities;
assuming this, the conditions in Equation           \eqref{wm} mean that all the $n$ fluids fulfill
the weak energy condition and at least one of them fulfills (as a strict inequality)
the strong energy condition (compare with Equations          \eqref{WEC} and \eqref{SEC}).
Let~us also remark that $w_i > (2/d) - 1 > -1$ if the $i$th fluid
is a radiation ($w_i = 1/d$) in spatial dimension $d \geqs 2$,
or a dust ($w_i = 0$) in spatial dimension $d \geqs 3$.
\vspace{0.15cm}
\item[({vi})] There exists $\tp \in (\ti, \tf)$ such that
\begin{equation}\label{tau1}
\dot{a}(\tau) > 0 \quad \mbox{for all\, $\tau \in (\ti,\tp)$}\,.
\end{equation}
\end{enumerate}

\begin{Proposition}\label{prop}
Under the assumptions $(i)$--$(vi)$, there exists a particle horizon at all times:
\begin{equation}\label{claim}
\Theta(\tau) < +\infty \quad \mbox{for all\; $\tau \in(\ti, \tf)$}\,.
\end{equation}
\end{Proposition}
\begin{proof}
Due to the alternative stated in Equation           \eqref{altern}, it suffices to prove that
\begin{equation}\label{sufprove}
\Theta(\tp) < + \infty\,,
\end{equation}
with $\tp$ the instant mentioned in Assumption (iv); hereafter, we show how
to derive Equation           \eqref{sufprove}.

Let us first consider the identity 
\eqref{Fid2};
multiplying both sides of this identity by the Hubble ratio $\dot{a}/a$
and taking into account that we are assuming $\sigma$ = +1,
by elementary computations, we obtain
\begin{equation}\label{proof1}
{\dot{a} \over a}\;\dot{\ff}^2 \,=\, {d-1 \over 2\,\kappa_d}\;
{d \over d\tau}\! \left[ -\,{\dot{a}^2 \over a^2} - {\kk \over a^2}
+ \sum_{i = 1}^n {\CC_i \over a^{d (1 + w_i)}} \right] .
\end{equation}

From here, recalling that $\dot{a}>0$ on $(\ti,\tp)${\footnote{Of course 
$\dot{\ff}^2\! \geqs 0$,
because we are considering a real scalar field.}}
we infer
\begin{equation}\label{57}
{d \over d\tau}\! \left[{\dot{a}^2 \over a^2} + {\kk \over a^2}
- \sum_{i = 1}^n {\CC_i \over a^{d (1 + w_i)}} \right] \leqs\, 0\,
\qquad \mbox{on\, $(\ti,\tp)$}\,.
\end{equation}

Thus, the function of time between the square brackets in Equation           \eqref{57}
is non-increasing on $(\ti,\tp]$. Due to this, the value of the function
at any instant $\tau$ in this interval is greater than or equal to
its value at $\tp$, a fact which can be expressed as
\begin{gather}
{\dot{a}^2(\tau) \over a^2(\tau)} \,\geqs\, \UU(a(\tau)) + \Ep \qquad
\mbox{for\, $\tau \in (\ti,\tp]$}\,; \label{proof2} \\
\UU(a) := -\, {\kk \over a^2} + \sum_{i = 1}^n {\CC_i \over a^{d (1 + w_i)}} \;, \qquad
\Ep := \left[{\dot{a}^2 \over a^2} + {\kk \over a^2}
- \sum_{i = 1}^n {\CC_i \over a^{d (1 + w_i)}}\right]_{\tau \,=\, \tp}\!. \label{UUEp}
\end{gather}

Incidentally, it should be noticed that Equation           \eqref{tau1} grants
$a(\tau) \!<\! a(\tp)$ for $\tau \!\in\! (\ti,\tp)$;
together with the assumption $\kk \leqs 0$ of Equation~\eqref{kneg}
and the conditions on $w_i, \CC_i$ stated in Equations~\eqref{wm}~and \eqref{Ci},
this yields
\begin{equation}\label{taka}
\UU(a(\tau)) + \Ep > 0 \qquad \mbox{for\, $\tau \in (\ti,\tp)$}\,.
\end{equation}

To proceed, notice that Equations          \eqref{proof2} and  \eqref{taka} imply
\begin{equation}\label{proof3}
{1 \over a(\tau)} \leqs {\dot{a}(\tau) \over a^2(\tau) \sqrt{\UU(a(\tau)) + \Ep}}
\qquad \mbox{for\, $\tau \in (\ti,\tp)$}\,.
\end{equation}

Using the above relation and recalling the definition 
\eqref{Theta}, we obtain
\begin{equation}\label{chain}
\Theta(\tp) \,= \int_{\ti}^{\tp}\! {d\tau \over a(\tau)}\,
\leqs \int_{\ti}^{\tp}\!
{\dot{a}(\tau)\; d\tau \over a^2(\tau)\, \sqrt{\UU(a(\tau))\! +\! \Ep}}\,
= \int_{0}^{a(\tp)}\!\! {d a \over a^2 \sqrt{\UU(a)\! +\! \Ep}}\;.
\end{equation} 

On the other hand, in view of Assumption  (v), we have
\begin{equation}\label{asy}
{1 \over a^2 \sqrt{\UU(a)\! +\! \Ep}}
\,=\, {1 \over \Qm^{\,1/2}}\;a^{{d(1 + \wm) \over 2} - 2}\;\big(1 + o(1)\big)
\qquad \mbox{for\; $a \to 0$}\,,
\end{equation}
where 
\begin{equation}
\wm := \max_{i \in \{1,...,n\}} w_i\,, \qquad
\Qm \;:= \!\!\!\sum_{\begin{array}{c}
\vspace{-0.6cm}\\
\mbox{{\scriptsize $i \in \{1,...,n\}$}} \vspace{-0.17cm}\\
\mbox{{\scriptsize s.t. $w_i = \wm$}}
\end{array}}\hspace{-0.25cm} \CC_i \,;
\end{equation}
note that the assumptions 
\eqref{Ci} and \eqref{wm} ensure
$\Qm > 0$ and $\wm > (2/d) -1$. \linebreak 
Since~$d(1\!+\!\wm)/2 - 2 > -1$,
the asymptotics 
\eqref{asy} indicates that
\begin{equation}\label{Intfin}
\int_{0}^{a(\tp)}\!\! {da \over a^2 \sqrt{\UU(a)\! +\! \Ep}} \,<\, +\infty\,.
\end{equation}

Summing up, Equations          \eqref{sufprove}, \eqref{chain} and \eqref{Intfin} yield the thesis 
\eqref{claim}.
\end{proof}

\section{Some Examples Where a Phantom Gives No Particle Horizon}\label{secPhan}
Contrary to the result for cosmological models with a quintessence
stated in Proposition~\ref{prop}, it~turns out
that the presence of a phantom often allows   disposing of the horizon problem.
To~support this claim, in the sequel, we present, as examples,
some models of the type described in Section~\ref{secCos} with a phantom
and no particle horizon.
Accordingly, throughout the present section, we~assume
\begin{equation}\label{sim}
\sigma = -1\,.
\end{equation}

Since we just want to ascertain the absence of a particle
horizon, we   produce very simple models; these could
be made more realistic via appropriate refinements 
that we prefer to postpone to future investigations. One
of the present simplifying assumptions is the existence
of just one type of perfect fluid ($n = 1$); 
for brevity, we denote the corresponding density, pressure
and parameters (see Equations          \eqref{wi}, \eqref{rhop} and \eqref{eqred}) with
\begin{equation}\label{par1}
\rho \equiv \rho_1\,, \qquad p \equiv p_1\,, \qquad
w \equiv w_1\,, \qquad \CC \equiv \CC_1\,.
\end{equation}

All the models described in the sequel are built using
the following strategy: first, prescribe the time dependence
of the scale factor $a(\tau)$; then, use the Einstein equations
to derive the time dependence of the scalar field $\phi(\tau)$
and the functional form of its potential $V(\phi)$.
This idea can be viewed as a special case of a general
``inverse approach'' to homogeneous and isotropic cosmological
models with a scalar field~\cite{Ba1,BaPa,ElMa,Lucc}, where the Einstein
equations are used to determine the field and its potential
after prescribing the time behavior of some relevant functions;
such an inverse approach was found to provide enticing results.

Let us go into the details of the above mentioned strategy,
which   is employed to construct the examples presented in
Sections~\ref{subs1}--\ref{subs3}. This relies on the following steps:

\begin{enumerate}[leftmargin=2.3em,labelsep=4mm]
\item[(a)] We choose a smooth function $\tau \!\in\! (\ti, \tf) \mapsto a(\tau)>0$,
where $-\infty \!\leqs\! \ti \!<\! \tf \!\leqs\! + \infty$;
we also prescribe the spatial curvature $\kk$, the coefficient $w$ in the equation
of state for the perfect fluid, and the constant $\CC$ determining its density.
These choices must comply with two basic conditions. Firstly, it is required that
\begin{equation}
a(\tau) \to 0 \quad \mbox{for\, $\tau \to \ti$}\,, \qquad
\Theta(\tau) := \!\int_{\ti}^\tau {d \tau' \over a(\tau')} = + \infty
\quad \mbox{for all\, $\tau\!\in\! (\ti, \tf)$}
\end{equation}
(Big Bang with no particle horizon). Secondly, it is required that
\begin{equation}\label{defchi}
\chi(\tau) \!>\! 0 \quad \mbox{for\, $\tau\!\in\!(\ti, \tf)$}\,, \qquad
\chi := {d-1 \over \kappa_d}
\left[-\,{\dot{a}^2 \over a^2} + {\ddot{a} \over a} - {\kk \over a^2}
+ {d\,(1 \!+\! w)\,\CC \over 2\,a^{d(1 + w)}}\right] ;
\end{equation}
note that $\chi$ is the right-hand side of Equation           \eqref{Fid2} (with $\sigma=-1$).
\vspace{0.15cm}
\item[(b)] After fulfilling Item (a), we set
\begin{equation}\label{deffi}
\phi(\tau) \,:=\, \fm + \!\int_{\ts}^{\tau}\! d \tau' \sqrt{\chi(\tau')}
\qquad \mbox{for\, $\tau \!\in\! (\ti, \tf)$}\,,
\end{equation}
where $\ts$ is arbitrarily fixed in the above interval, and $\fm$
is an arbitrary constant in $(-\infty,+\infty)$.
It should be noticed that the choice of $\ts$ and $\fm$ is immaterial,
since any change of this quantities is equivalent to shifting the scalar field
by a physically irrelevant additive constant; taking this into account,
in the following examples, we   fix $\ts$ and $\fm$     to simplify
the expression of $\phi(\tau)$.
In any case, the function $\tau \mapsto \phi(\tau)$ is smooth
with derivative
\begin{equation}\label{derivfi}
\dot{\phi}(\tau) = \sqrt{\chi(\tau)}>0
\end{equation}
and Equation           \eqref{Fid2}
is fulfilled by construction. Being monotonic, the function
$\tau \to \phi(\tau)$ is one to one between $(\ti, \tf)$
and a suitable interval $(\fin, \ffi)$, with a smooth inverse function
\begin{equation}
\ff \in (\fin, \ffi) \,\mapsto\, \tau(\ff) \in (\ti, \tf)\,.
\vspace{0.cm}
\end{equation}
\item[(c)] Finally, we put
\begin{gather}
W \!:= {d\!-\!1 \over 2\,\kappa_d}\!
\left[(d\!-\!1)\,{\dot{a}^2 \over a^2} + {\ddot{a} \over a} + (d\!-\!1)\,{\kk \over a^2}
- {d\,(1\! -\! w)\,\CC \over 2\,a^{d (1 + w)}}\right] \!:
(\ti, \tf) \to (-\infty,+\infty)\,; \label{defizw}\\
V \!: (\fin, \ffi) \to (-\infty,+\infty)\,, \qquad
\ff \mapsto V(\ff) := W(\tau(\ff))\,. \label{defizv}
\end{gather}

Clearly, $W$ is the left-hand side of Equation           \eqref{Vf}; thus, the cited equation
holds true by construction if we define the field potential $V(\ff)$
as in Equation           \eqref{defizv}.
\end{enumerate}

Summing up, with our algorithm we fulfill on the interval $(\ti, \tf)$
both Equations          \eqref{Fid2} and~\eqref{Vf}, which are equivalent to the Einstein
equations 
\eqref{E1} and \eqref{E2}.
Since $\dot{\phi}$ never vanishes on $(\ti, \tf)$, the Klein--Gordon
equation 
\eqref{ffeq} is granted to hold as well on this interval
(recall the identity 
\eqref{ident}).

In the applications of the strategy Items (a)--(c)     presented in
Sections~\ref{subs1}--\ref{subs3}, we have $\ti = - \infty$ or $\ti = 0$. 
In all these examples, $\tf = + \infty$ and, for $\tau \to + \infty$, 
$a(\tau)$ diverges exponentially while $\rho(\tau), p(\tau)$ vanish
and $\rf(\tau), \pf(\tau)$ approach finite values;
thus, neither a Big Rip nor a Little Rip occurs.
Let us also remark that, in the same limit $\tau \to +\infty$,
the equation of state parameter $\wf$ for the field approaches
$-1$ from below.

{Before proceeding,     to avoid misunderstandings, let us point
out that all the models to be described in the sequel depend (apart
from a dimensionless parameter $A$) on a positive constant $\HH$,
with the dimensions of an inverse time. This constant is strictly related
to, but does not necessarily coincide with the Hubble ratio $H := \dot{a}/a$~{\footnote{In fact, the equality $\dot{a}(\tau)/a(\tau) = \HH$
(for any $\tau \in (-\infty,+\infty)$)
holds true only for the model described in       Section~\ref{subs1}.
In~the cases discussed in Sections~\ref{subs2} and~\ref{subs3}, one has,
respectively, $\dot{a}(\tau)/a(\tau) \to \HH$ for $\tau \to + \infty$
and $\dot{a}(\tau)/a(\tau) \to 2 \HH$ in the same limit.}}.}

\subsection{A de Sitter Cosmology with Zero Spatial Curvature}\label{subs1}
Let us assume that the scale factor is that of a de Sitter geometry;
namely, we put
\begin{equation}\label{adS}
a(\tau) := A\,e^{\HH \tau} \quad
\mbox{for $\tau \in (-\infty,+\infty)$\; (with $A,\HH > 0$ constants)} \,.
\end{equation}

Notice that $\dot{a}(\tau) > 0$ and $\ddot{a}(\tau) > 0$ (accelerated expansion) for all
$\tau \in (-\infty,+\infty)$.

We       apply the previous scheme Items (a)-(c) with $\ti = -\infty$,
$\tf = + \infty$ and the above choice 
\eqref{adS} of $a$. Of course,
this setting gives a Big Bang in the infinitely remote past, with no particle horizon:
\begin{equation}
a(\tau) \to 0 \quad \mbox{for\, $\tau \to - \infty$}\,; \qquad
\Theta(\tau) = \!\int_{-\infty}^{\tau} {d \tau' \over a(\tau')} \,=\, + \infty\,.
\end{equation}

Concerning the spatial curvature and the matter fluid, we assume
\begin{gather}
\kk = 0\,; \label{k0}\\
w > -1 \quad\; \mbox{and} \quad\; \CC > 0 \,.
\end{gather}

In the case under analysis, the function $\chi$ defined by Equation           \eqref{defchi}
is such that $\chi(\tau) = d (d-1)$ $(1 + w) \CC/(2 \kappa_d A^{d (1 + w)}) \times e^{- d (1+ w) \HH \tau}>0$
for all $\tau \in (-\infty,+\infty)$;
thus, Equation           \eqref{deffi} with $\fm = 0$ and $\ts \to + \infty$
(see the comments in Item (b) of the general strategy) gives
\begin{equation}\label{fdS}
\ff(\tau) =
-\,{1 \over A^{d(1 + w)/2}\, \HH}\,\sqrt{{2\,(d-1)\,\CC \over d\,\kappa_d\, (1 + w)}}\;
e^{- d (1 + w) \HH \tau/2} \qquad \mbox{for\, $\tau \in (-\infty,+\infty)$}\,.
\end{equation}

The function $\tau \mapsto \ff(\tau)$ maps $(-\infty,+\infty)$
to $(-\infty,0)$ and is strictly increasing; the~corresponding
inverse function is described by the relation
\begin{equation}\label{tauf}
\tau(\ff) =
-\,{2 \over d (1\!+\!w) \HH}\,\log\!\left[A^{d(1+w)/2}\, \HH\,
\sqrt{d\,\kappa_d\,(1 + w) \over 2(d-1)\CC}\,(-\,\ff)\right]
\quad \mbox{for $\ff < 0$}\,.
\end{equation}

To go on, we note that Equation           \eqref{defizw} gives, in the present case,
$W(\tau) = d(d\!-\!1)\HH^2\!/(2\kappa_d)
- d(d\!-\!1)$ $(1\!-\!w) \CC/(4 \kappa_d A^{d(1+w)}) \times e^{-d(1+w)\HH \tau}$;
from here and   Equations          \eqref{defizv} and \eqref{tauf}, we obtain for the field potential
the expression
\begin{equation}\label{VdS}
V(\ff) = -\,{d^2 (1 \!-\! w^2) \HH^2 \over 8}\; \ff^2
+\, {d\,(d-1) \HH^2 \over 2\,\kappa_d}\,.
\end{equation}

It should be noticed that, in principle, Equation           \eqref{VdS} holds true
for $\ff < 0$; however, it can be used to define $V$
on the whole interval $(-\infty,+\infty)$.

 Equation  \eqref{VdS} makes patent that $V(\ff)$ is not bounded from
below when $|w| < 1$; contrary to the case of a quintessence, this is not
a problematic feature in the case of a phantom, since the latter naturally
evolves towards maxima of the potential (see the comments after Equation           \eqref{KG}).
In fact, both $\ff(\tau)$ and $V(\ff(\tau)) \equiv  W(\tau)$
diverge to a negative infinity for $\tau \to -\infty$ (i.e., close to the Big Bang),
while we have
\begin{gather}
\ff(\tau) \to 0\,, \quad V(\ff(\tau)) \to \max V \!= V(0) = {d\,(d-1) \HH^2 \over 2\,\kappa_d}
\qquad \mbox{for $\tau \to +\infty$}\,.
\end{gather}

The above relations and Equations          \eqref{wi}, \eqref{eqred} and \eqref{rhofpf} yield
\begin{gather}
\rho(\tau) \to 0\,, \quad p(\tau) \to 0\,, \quad 
\rf(\tau) \to {d (d\!-\!1) \HH^2 \over 2\,\kappa_d}\,, \quad
\pf(\tau) \to -\,{d (d\!-\!1) \HH^2 \over 2\,\kappa_d} \qquad 
\mbox{for $\tau \!\to\! +\infty$}\,.
\end{gather}

Finally, let us remark that Equations          \eqref{wff}, \eqref{Fid2}, \eqref{Vf} and \eqref{adS} imply
\begin{equation}
\wf(\tau) = \left\{\!\!\begin{array}{ll}
\dd{w + O\big(e^{d(1+w)\HH \tau}\big)} & \dd{\mbox{for $\tau \to -\infty$}\,,} \vspace{0.1cm} \\
\dd{-1 - {(1\!+\!w)\,\CC \over A^{d(1+w)}\,\HH^2}\,e^{-d(1+w)\HH \tau}\!
+ O\big(e^{-2d(1+w)\HH \tau}\big)} & \dd{\mbox{for $\tau \to +\infty$}\,.}
\end{array}\right.
\end{equation}

The above relations show that the phantom $\ff$ behaves as
a perfect fluid with equation of state parameter approaching $w$ near
the Big Bang ($\tau \to -\infty$), while it behaves as a cosmological
constant for large times ($\tau \to +\infty$; see the comments
at the end of Section~\ref{secCos}).

Before proceeding, we would recall that a model similar to the one analyzed
in the present section is   considered in~\cite{BaPa}.
In fact, on Page 10 of the cited work, the authors mentioned the case
of a spatially flat de Sitter cosmology, including a perfect fluid and
a self-interacting scalar field; correspondingly, they report a quadratic
expression (comparable to the one in Equation           \eqref{VdS}) for the field potential.
This work never mentions phantom fields explicitly, but the
appearance of square roots of negative argument in certain equations
for scalar fields can be re-interpreted making reference to the phantom case.
In the specific de Sitter model considered in~\cite{BaPa},
a square root of negative argument actually appears if one uses
Equation           (23) therein, with $F'(\omega) = 0$ and $\gamma, \rho_{m0} > 0$.

\subsection{A Model with Big Bang at Finite Cosmic Time and Negative Curvature}\label{subs2}
Let us consider the scale factor
\begin{equation}\label{a2}
a(\tau) := A\,\sinh(\HH\, \tau) \quad
\mbox{for $\tau \in (0,+\infty)$\; (with $A,\HH > 0$ constants)\,, }
\end{equation}
noting that $\dot{a}(\tau) > 0$ and $\ddot{a}(\tau) > 0$
for all $\tau \!\in\! (0,+\infty)$.

In the following, we relate the scale factor 
\eqref{a2} to the general
scheme (Items (a)--(c)) with $\ti = 0$ and $\tf = + \infty$.
Since $a(\tau) = A \HH \tau + O(\tau^3)$ for $\tau \to 0$, we have a Big Bang
at time zero, and the integral $\Theta(\tau)$ of Equation           \eqref{Theta} diverges logarithmically;
thus, there is no particle horizon, regardless of the spatial curvature $\kk$.

From here to the end of the present subsection, we focus on the physically relevant case
of a four-dimensional spacetime, choosing for the spatial curvature a special value
that greatly simplifies the analysis of this model; more precisely, we assume
\begin{equation}
d = 3 \qquad \mbox{and} \qquad \kk = -\, A^2\, \HH^2 <\, 0\,.
\end{equation}

The perfect fluid is assumed to be of radiation type:
\begin{equation}
w = 1/3 \quad\; \mbox{and} \quad\;  \CC > 0\,.
\end{equation}

In this case, the function $\chi$ defined by Equation           \eqref{defchi}
is such that $\chi(\tau) = 4 \CC/(\kappa_3 A^4) \times \sinh^{-4}(\HH\, \tau) > 0$
for all $\tau \in (0,+\infty)$. Thus, Equation           \eqref{deffi} with
$\fm = -\,2 \sqrt{\CC/\kappa_3}/(A^2\, \HH)$ and $\ts \to +\infty$
(see again the comments in Item (b) of the general strategy) gives
\begin{equation}\label{f2}
\ff(\tau) = -\, {2\sqrt{\CC/\kappa_3} \over A^2\, \HH}\; \mbox{coth}(\HH\, \tau) \,,
\end{equation}
where $\mbox{coth}y := \cosh y/\sinh y$ indicates the hyperbolic cotangent.
The function $\tau \to \phi(\tau)$ maps $(0,+\infty)$
to $(-\infty, - 2 \sqrt{\CC/\kappa_3}/(A^2 \HH))$ and is strictly increasing,
with inverse
\begin{equation}\label{t2}
\tau(\ff) = {1 \over \HH}\; \mbox{acoth}\!
\left[{A^2\, \HH \over 2 \sqrt{\CC/\kappa_3}}\; (-\,\ff) \right]
\qquad \mbox{for}\; \ff < -\,{2 \sqrt{\CC/\kappa_3} \over A^2\, \HH}\; ;
\end{equation}
where ``$\mbox{acoth}$'' indicates the inverse of the hyperbolic cotangent.
The next step relies on Equation~\eqref{defizw}, giving
in the present case $W(\tau) = (3 \HH^2\!/\kappa_3) - \CC/(\kappa_3 A^4) \times \sinh^{-4}(\HH \tau)$;
from~here and   Equations          \eqref{defizv} and \eqref{tauf},
using the basic identity $\sinh( \mbox{acoth}x) = 1/\sqrt{x^2\!-\!1}$,
we obtain for the field potential the expression
\begin{equation}\label{V2}
V(\ff) = - \,{A^4 \HH^4 \kappa_3 \over 16\,\CC}\; \ff^4
+ {\HH^2\! \over 2}\; \ff^2
+ {\HH^2\! \over \kappa_3}\!\left(3 - {\CC \over A^4 \HH^2}\right),
\end{equation}
which in fact makes sense for all $\ff \in (-\infty,+\infty)$.
In passing, let us notice that the potential 
\eqref{V2} is
of Higgs  type, with the opposite sign (as typical for a phantom).

The above results show that both $\ff(\tau)$ and $V(\ff(\tau))$ diverge to
negative infinities in the Big Bang limit $\tau \to 0$, while we have
\begin{gather}
\ff(\tau) \!\to\! - \,{2\,\sqrt{\CC/\kappa_3} \over A^2\, \HH}\,, \;\;
V(\ff(\tau)) \!\to\! \max V \!= V\!\left(\!- \,{2\,\sqrt{\CC/\kappa_3} \over A^2\, \HH}\right)\!
= {3 \HH^2 \over \kappa_3}
\quad\; \mbox{for $\tau \!\to\! +\infty$}\,.
\end{gather}

On account of the above relations and   Equations          \eqref{wi}, \eqref{eqred} and \eqref{rhofpf}, we get
\begin{gather}
\rho(\tau) \to 0\,, \quad p(\tau) \to 0\,, \quad 
\rf(\tau) \to {3 \HH^2 \over \kappa_3}\,, \quad
\pf(\tau) \to -\,{3 \HH^2 \over \kappa_3} \qquad 
\mbox{for $\tau \!\to\! +\infty$}\,.
\end{gather}

To conclude, let us mention that Equations          \eqref{wff}, \eqref{Fid2}, \eqref{Vf}
and \eqref{a2} give
\begin{equation}\label{wf2}
\wf(\tau) = \left\{\!\!\begin{array}{ll}
\dd{{1 \over 3} + O\big((\HH \tau)^2\big)} & \dd{\mbox{for $\tau \to 0$}\,,} \vspace{0.1cm} \\
\dd{-1 - {64\,\CC \over 3 A^4 \HH^2}\,e^{-4 \HH \tau}\!
+ O\big(e^{-6 \HH \tau}\big)} & \dd{\mbox{for $\tau \to +\infty$}\,.}
\end{array}\right.
\end{equation}

Similar  to the de Sitter model discussed in       Section~\ref{subs1},
the above relations allow us to infer that the phantom $\ff$ approaches
a perfect fluid of radiation type near the Big Bang,
while it mimics a cosmological constant for large times.

\subsection{A Model with Big Bang at Finite Cosmic Time and Zero Curvature}\label{subs3}
In this last example, we investigate the cosmological model corresponding
to the scale factor
\begin{equation}\label{a3}
a(\tau) = A\;{\sinh^2(\HH \tau) \over \HH \tau} \quad
\mbox{for $\tau \in (0,+\infty)$\; (with $A,\HH > 0$ constants)\,.}
\end{equation}

One checks that, even in this case, it is
$\dot{a}(\tau)\!> \!0$ and $\ddot{a}(\tau) \!>\! 0$ for all
$\tau \!\in\! (0,+\infty)$.

In the sequel, we implement the general strategy Items (a)--(c)
with $\ti = 0$ and $\tf = + \infty$.
Since~$a(\tau) = A \HH \tau + O(\tau^3)$ for $\tau \to 0$,
a Big Bang occurs at time zero and the integral in Equation~\eqref{Theta}
diverges logarithmically for any $\tau > 0$; consequently, there is no
particle horizon for any value of the spatial curvature $\kk$.

We restrict the attention to a spatially flat
four-dimensional universe, filled with radiation. Accordingly, we set
\begin{gather}
d =3\,, \qquad \kk = 0\,, \\
w = 1/3 \,, \qquad  \CC > 0\,; \label{wqq}
\end{gather}
the assumption on the constant $\CC$ is provisional,
and will be strengthened shortly afterwards.

In the present framework, Equation           \eqref{defchi} gives
\begin{equation}\label{f3}
\chi(\tau) =
{\HH^2 \over \kappa_3}\! \left[{2 \over t^2} - {4 \over \sinh^2(t)}
+ {4\,\Om\;t^4 \over \sinh^8(t)}\right]_{t \,=\, \HH \tau} ;
\end{equation}
this equation involves the dimensionless time coordinate
$t := \HH \tau \!\in\! (0,+\infty)$
and the dimensionless density parameter
\begin{equation}
\Om := {\CC \over A^4 \HH^2} \,>\, 0\,.
\end{equation}

It should be noticed that the condition $\chi(\tau) >0$
for all $\tau \in (0,+\infty)$ is fulfilled if and only if
\begin{equation}\label{Omc}
\Om > \Om_c\,, \qquad
\Om_c :=\! \max_{t \,\in\, (0,+\infty)}\! \left[ {\sinh^8(t) \over 2\;t^4}
\left({2 \over \sinh^2(t)} - {1 \over t^2}\right)\right]\! = 1.24318...\;.
\end{equation}

From here to the end of the subsection, we enforce the
condition 
\eqref{wqq} on $\CC$ assuming that Equation \eqref{Omc}   holds.
Having granted the positivity of $\chi$, we proceed to define
a function $\tau \mapsto \ff(\tau)$ using the relation 
\eqref{deffi} with $\fm = 0$ and $\ts = 1/\HH$ (see again the comments in Item (b)
of the general strategy); this gives
\begin{equation}\label{fExp}
\ff(\tau) =
{1 \over \sqrt{\kappa_3}} \int_{1}^{\HH \tau}\!\! dt\,
\sqrt{{2 \over t^2} - {4 \over \sinh^2(t)} + {4\,\Om\;t^4 \over \sinh^8(t)}}
\qquad \mbox{for $\tau \in (0,+\infty)$}\,.
\end{equation}

By construction, this function is monotonically increasing;
besides, using Equation           \eqref{fExp}, one~can show that $\phi(\tau) \to - \infty$
for $\tau \to 0$ and $\phi(\tau) \to + \infty$ for $\tau \to + \infty$
(more details on these issues are given in the sequel). Thus, the map
$\tau \mapsto \ff(\tau)$ is one to one between $(0,+\infty)$ and $(-\infty,+\infty)$
and possesses a smooth inverse
\begin{equation}\label{invf}
\ff \in (-\infty,+\infty) \,\mapsto\, \tau(\ff) \in (0,+\infty) \,.
\end{equation}

Unfortunately, we cannot give simple analytic expressions for
the maps 
\eqref{fExp} and~\eqref{invf}; the same problem will therefore affect
the field potential $V$. The latter is determined by the prescriptions 
\eqref{defizw} and \eqref{defizv}, which, in the present case, give
\begin{gather}
W(\tau) = {\HH^2 \over \kappa_3}\! \left[2 + {4 \over t^2}
- {12\,\cosh(t) \over t\,\sinh(t)} + {10\,\cosh^2(t) \over \sinh^2(t)}
- {\Om\;t^4 \over \sinh^8(t)} \right]_{t\,=\,\HH \tau} \quad \mbox{for $\tau \in (0,+\infty)$}\,,\label{W3} \\
V(\ff) = W(\tau(\ff)) \qquad \mbox{for $\ff \in (-\infty,\infty)$}\,. \label{V3}
\end{gather}

\begin{figure}[H]
    \centering
        \begin{subfigure}[b]{0.46\textwidth}
                \includegraphics[width=\textwidth]{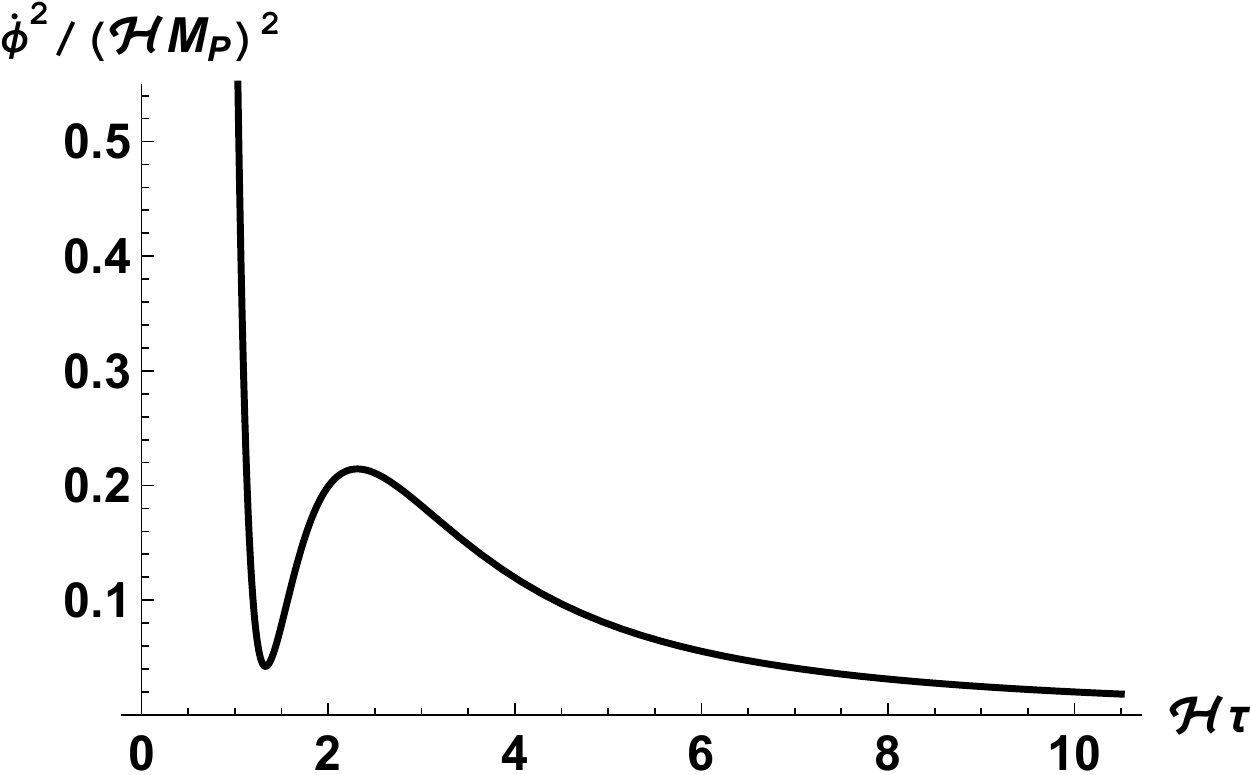}
                \caption*{(\textbf{a}) }
        \end{subfigure}
        \hspace{0.4cm}
        \begin{subfigure}[b]{0.46\textwidth}
                \includegraphics[width=\textwidth]{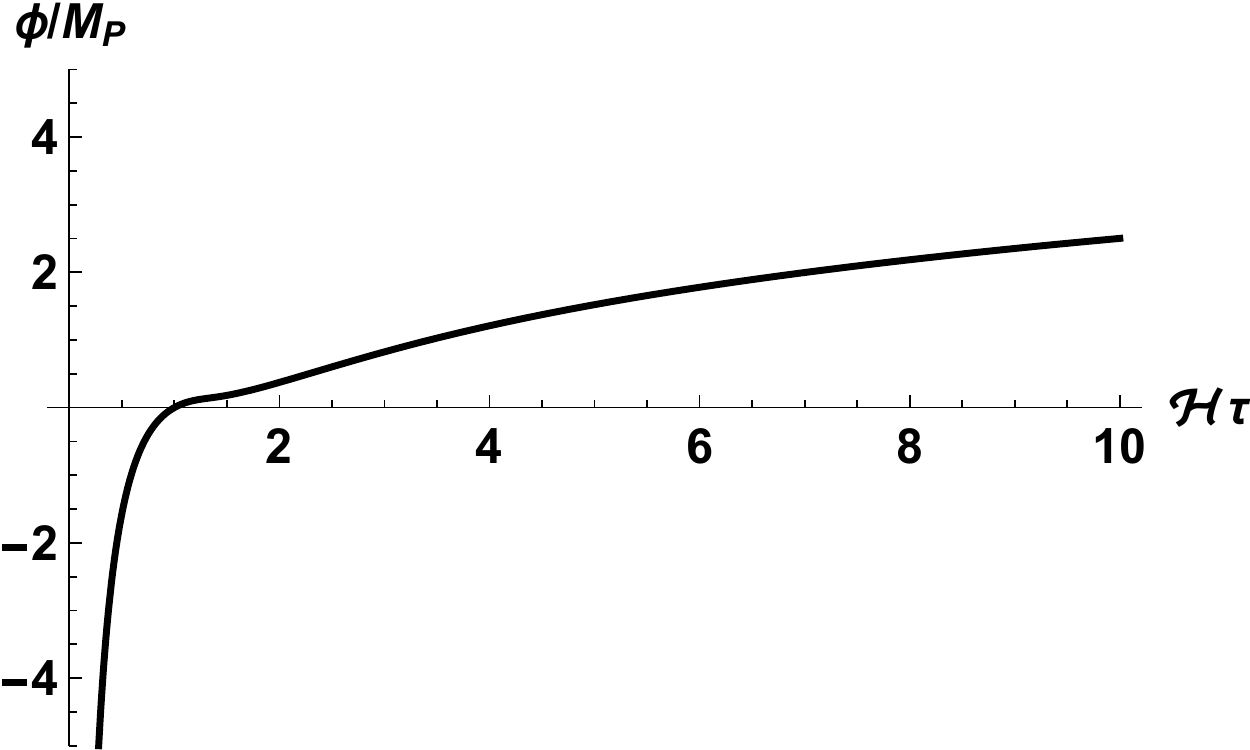}
                \caption*{(\textbf{b}) }
        \end{subfigure}
        \vspace{6pt}
\caption{(\textbf{a}) Plot of the map $\tau \mapsto \dot{\ff}^2(\tau)/(\HH \ma)^2$ for $\Om = 1.5$; (\textbf{b}) Plot of the map $\tau \mapsto \ff(\tau)/\ma$ for $\Om = 1.5$.} 
\label{fig:Fi}
\end{figure}

\begin{figure}[H]
    \centering
        \begin{subfigure}[b]{0.46\textwidth}
                \includegraphics[width=\textwidth]{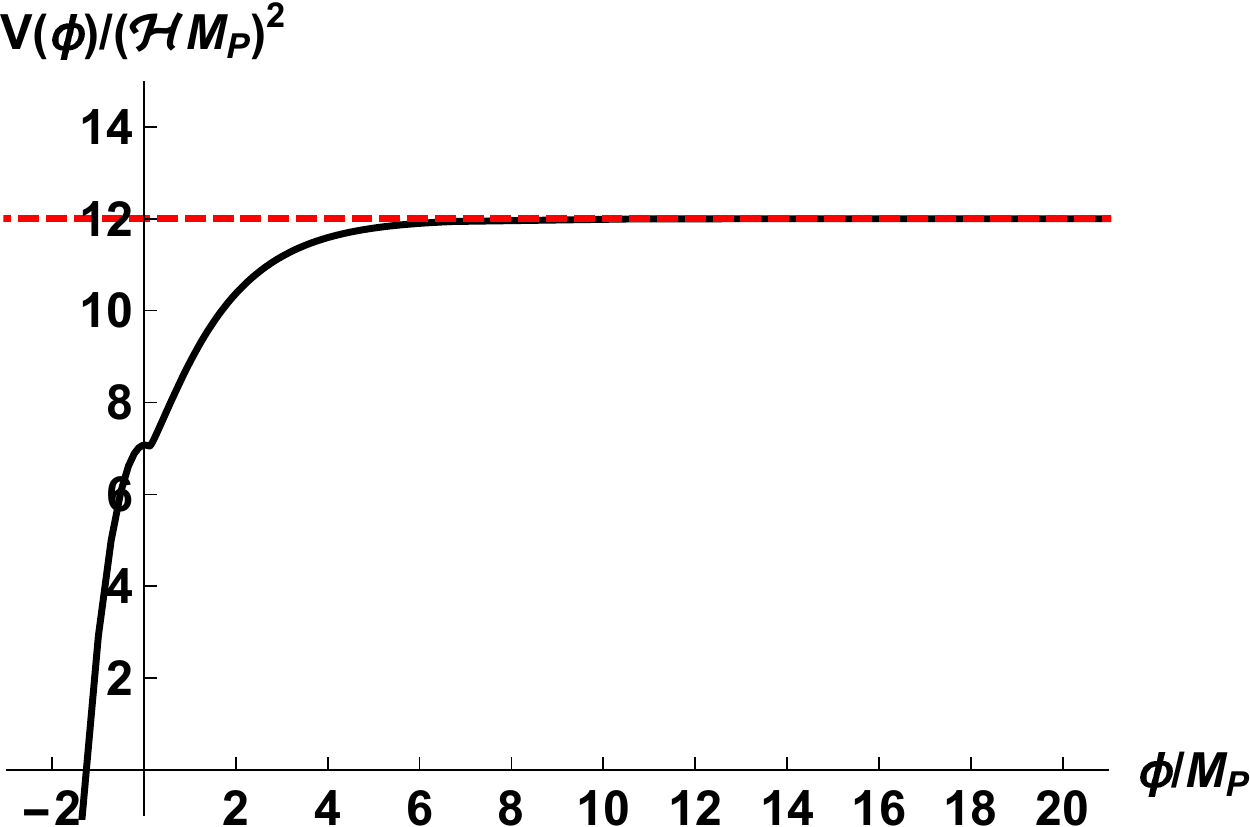}
                \caption*{(\textbf{a}) }
        \end{subfigure}
        \hspace{0.4cm}
        \begin{subfigure}[b]{0.46\textwidth}
                \includegraphics[width=\textwidth]{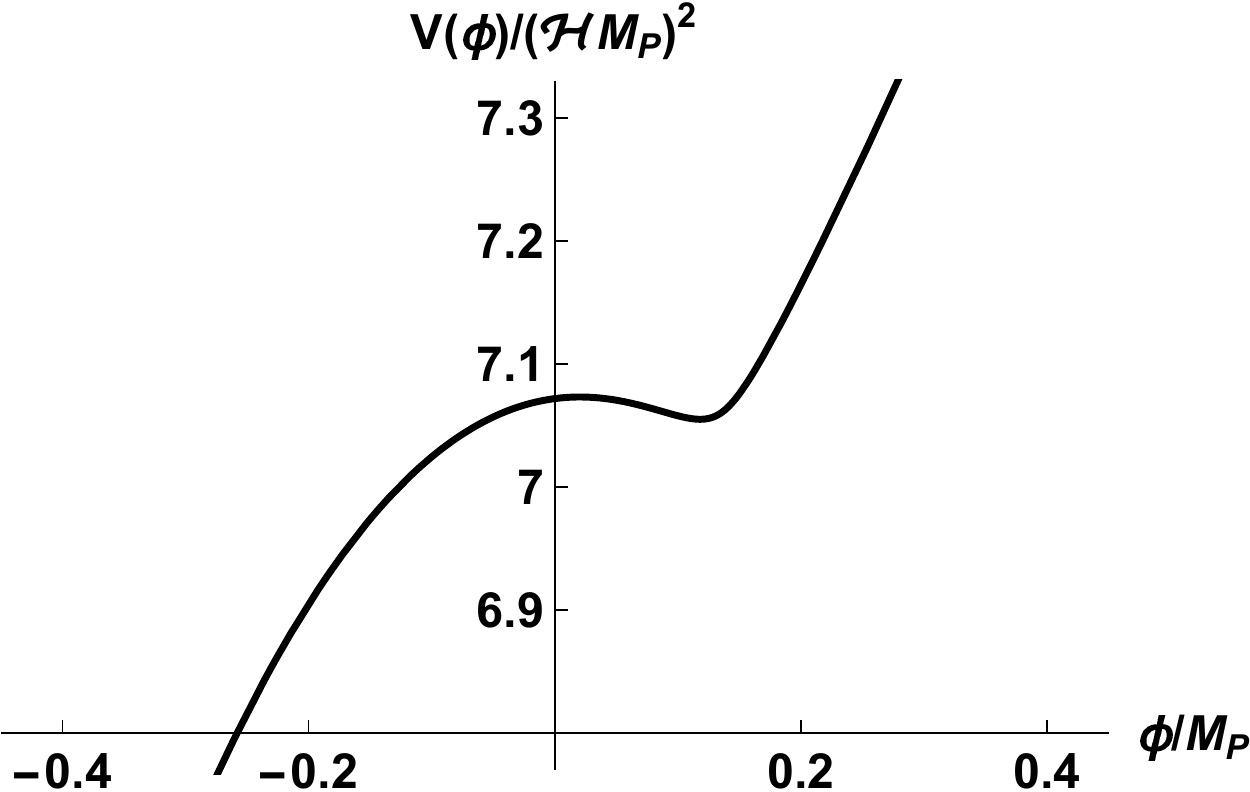}
                \caption*{(\textbf{b}) }
        \end{subfigure}
        \vspace{6pt}
\caption{(\textbf{a}) Graph of $V(\ff)/(\HH \ma)^2$ as a function of $\ff/\ma$, for $\Om = 1.5$
				(obtained by parametric plotting of the curve
                $\tau \!\mapsto\! (\ff(\tau)/\ma, W(\tau)/(\HH \ma)^2)$; 
                (\textbf{b}) Zoom of (\textbf{a}), showing that
                $V(\ff)$ has a local minimum and a local maximum
                near $\ff = 0$ for $\Om = 1.5$.} 
\label{fig:VFi}
\end{figure}

The above functions are better understood introducing the Planck mass
\begin{equation}
\ma := {1 \over \sqrt{\kappa_3}}
\end{equation}
and considering the dimensionless ratios $\ff/\ma$, $V(\ff)/(\HH \ma)^2$
(recall that $c = 1$, $\hbar = 1$ in our units). {Figures~\ref{fig:Fi}a and~\ref{fig:VFi}b} refer to an
admissible value of $\Om > \Om_c$, namely $\Om = 1.5$. More precisely:

\begin{itemize}[leftmargin=2.3em,labelsep=5.5mm]
\item Figure~\ref{fig:Fi}a is a plot of the ratio $\dot{\ff}^2/(\HH \ma)^2$ (again,
a dimensionless quantity) as a function of $\HH \tau$; this function is
known explicitly due to Equations          \eqref{derivfi} and \eqref{f3}.
\vspace{0.15cm}
\item Figure~\ref{fig:Fi}b is a plot of $\ff/\ma$ as a function of $\HH \tau$: this
was obtained from Equation           \eqref{fExp} for $\ff(\tau)$, computing numerically
the integral appearing therein.
\vspace{0.15cm}
\item Figure~\ref{fig:VFi}a is a plot of $V(\ff)/(\HH \ma)^2$ as a function of $\ff/\ma$;
this was obtained as the curve with parametric representation
$\tau\mapsto (\ff(\tau)/\ma, W(\tau)/(\HH \ma)^2)$, using
for $W(\tau)$ the explicit expression 
\eqref{W3} and, again, computing
numerically $\ff(\tau)$. Figure~\ref{fig:VFi}b is a zoom of the same plot,
showing more clearly that the map $\ff/\ma \mapsto V(\ff)/(\HH \ma)^2$
has a local minimum and a local maximum near $\ff = 0$
(for the chosen value $\Om = 1.5$;
these local extremal points are not present for much larger values of $\Om$,
e.g., for $\Om \gtrsim 70$).
\end{itemize}

To conclude, let us spend a few words about the certain asymptotic
features of the model, which can be explicitly determined.

First, let us remark that Equations          \eqref{fExp} and \eqref{W3} imply{\footnote{To derive the asymptotic relation for $\ff(\tau)$ written
in Equation           \eqref{asy0}, it should be noticed that the integrand function on
the right-hand side of Equation           \eqref{fExp} fulfills
$$ \sqrt{{2 \over t^2} - {4 \over \sinh^2(t)} + {4\,\Om\;t^4 \over \sinh^8(t)}}
= {2\,\sqrt{\Om} \over t^2} + O(1) \qquad \mbox{for $t \to 0^+$} . $$}}
\begin{equation}\label{asy0}
{\ff(\tau) \over \ma} = -\,{2\,\sqrt{\Om} \over \HH \tau} + O(1)\,, \quad
{W(\tau) \over (\HH \ma)^2} = -\,{\Om \over (\HH \tau)^4} + O\!\left({1 \over (\HH \tau)^2}\right)
\quad\; \mbox{for $\tau \to 0$}\,;
\end{equation}
these facts and Equation           \eqref{V3} allow us to infer that
\begin{equation}\label{asyVF0}
{V(\ff) \over (\HH \ma)^2} = -\,{1 \over 16\,\Om}\! \left({\ff \over \ma}\right)^{\!\!4} +
O\!\left(\!\left({\ff \over \ma}\right)^{\!\!2}\right)
\qquad \mbox{for $\ff \to -\infty$}\,.
\end{equation}

\unskip

Secondly, let us return to Equations          \eqref{fExp} and \eqref{W3} and note that{\footnote{To derive the asymptotic relation for $\ff(\tau)$ written in Equation           \eqref{asyInf},
it should be noticed that the integrand function on the right-hand side of Equation           \eqref{fExp}
fulfills
$$ \sqrt{{2 \over t^2} - {4 \over \sinh^2(t)} + {4\,\Om\;t^4 \over \sinh^8(t)}}
= {\sqrt{2} \over t} + O(t\,e^{-2t}) \quad\; \mbox{for $t \to +\infty$}\,.$$}}
\begin{equation}\label{asyInf}
{\ff(\tau) \over \ma} = \sqrt{2}\, \log(\HH \tau) + O(1)\,, \quad
{W(\tau) \over (\HH \ma)^2} = 12 + O\!\left({1 \over \HH \tau}\right)
\quad\; \mbox{for $\tau \to +\infty$}\,;
\end{equation}
along with Equation           \eqref{V3}, this implies
\begin{equation}
{V(\ff) \over (\HH \ma)^2} = 12 + O\big(e^{-{\ff/(\sqrt{2} \ma)}}\big)
\qquad \mbox{for $\ff \to +\infty$}\,.
\end{equation}

Due to Equations          \eqref{wi}, \eqref{eqred}, \eqref{rhofpf} and \eqref{derivfi},
the above relations imply
\begin{gather}
\rho(\tau) \to 0\,, \quad p(\tau) \to 0\,, \quad 
\rf(\tau) \to 12\,(\HH \ma)^2\,, \quad
\pf(\tau) \to -\,12\,(\HH \ma)^2 \quad\;
\mbox{for $\tau \!\to\! +\infty$}\,.
\end{gather}

In conclusion, let us point out that Equations          \eqref{wff}--\eqref{E2} and
\eqref{a3} yield an explicit expression for the coefficient $\wf(\tau)$ in the
field equation of state; this in turns implies the asymptotics
\begin{equation}\label{wasi}
\wf(\tau) = \left\{\!\!\begin{array}{ll}
\dd{{1 \over 3} + O\big((\HH \tau)^2\big)} & \dd{\mbox{for $\tau \to 0$}\,,} \vspace{0.1cm} \\
\dd{-1 - {1 \over 6\,(\HH \tau)^2} + O\!\left(\!{1 \over (\HH \tau)^3}\!\right)} & \dd{\mbox{for $\tau \to +\infty$}\,,}
\end{array}\right.
\end{equation}
which can be commented similarly to
Equation           \eqref{wf2} of the previous section.

\vspace{6pt} 



\authorcontributions{M.G. contributed mainly to aspects of the paper related to the general setting of scalar cosmologies; and D.F. and L.P. contributed mainly to the part of the paper related to Proposition~\ref{prop} and to the three cosmological models with a phantom field.}

\funding{This research was funded by: INdAM, Gruppo Nazio\-nale per la Fisica Matematica;
Istituto Nazionale di Fisica Nucleare; MIUR, PRIN 2010 Research Project
``Geometric and analytic theory of Hamiltonian systems in finite and infinite dimensions'';
Universit\`{a} degli Studi di Milano.}

\acknowledgments{
We acknowledge the referees for useful comments and bibliographical indications, which led to an essential improvement in the presentation of the results of this paper.
}

\conflictsofinterest{The authors declare no conflict of interest.} 



\reftitle{References}





\end{document}